\documentclass[%
 aip,
 amsmath,amssymb,
 reprint,%
]{revtex4-1}

\usepackage{graphicx}
\usepackage{dcolumn}
\usepackage{bm}

\usepackage[utf8]{inputenc}
\usepackage[T1]{fontenc}
\usepackage{mathptmx}

\begin{document}

\preprint{AIP/123-QED}

\title[Machine-Learning Many-Body Potentials for Colloidal Systems]{Machine-Learning Many-Body Potentials for Colloidal Systems}

\author{Gerardo Campos-Villalobos}
 \email{g.d.j.camposvillalobos@uu.nl}
\affiliation{ 
Soft Condensed Matter, Debye Institute for Nanomaterials Science, Utrecht University, Princetonplein 1, 3584 CC Utrecht, The Netherlands
}%
\author{Emanuele Boattini}%
\affiliation{ 
Soft Condensed Matter, Debye Institute for Nanomaterials Science, Utrecht University, Princetonplein 1, 3584 CC Utrecht, The Netherlands
}%

\author{Laura Filion}%
\affiliation{ 
Soft Condensed Matter, Debye Institute for Nanomaterials Science, Utrecht University, Princetonplein 1, 3584 CC Utrecht, The Netherlands
}%

\author{Marjolein Dijkstra}%
 \email{m.dijkstra@uu.nl}
\affiliation{ 
Soft Condensed Matter, Debye Institute for Nanomaterials Science, Utrecht University, Princetonplein 1, 3584 CC Utrecht, The Netherlands
}%

\date{\today}

\begin{abstract}
Simulations of colloidal suspensions  consisting of mesoscopic particles and  smaller species like ions or depletants  are computationally challenging as different length and time scales are involved. 
 Here, we introduce a machine learning (ML)  approach  in which  the degrees of freedom of the microscopic species are integrated out and the mesoscopic particles interact with effective many-body potentials, which we  fit   as a function of all colloid coordinates  with a set of symmetry functions. We apply this approach to a colloid-polymer mixture. Remarkably, the ML potentials can be assumed to be effectively state-independent and can be used in direct-coexistence simulations.   We show that our ML method reduces the computational cost by several orders of magnitude compared to a numerical evaluation, and accurately describes the phase behavior and structure, even for state points where the effective potential is largely determined by many-body contributions. 

\end{abstract}

\maketitle

\section{\label{sec:intro}Introduction}

Colloidal suspensions  consist of mesoscopic particles suspended in a solvent.  The effective interactions between the colloids can be modified by the addition of smaller species like ions, depletants, ligands, or polymers. For instance, the addition of non-adsorbing polymer to a colloidal suspension induces an  effective attraction between the colloids. This arises from an increase in configurational entropy of the polymer as colloidal particles approach each other and their depletion zones overlap as described by Asakura and Oosawa in 1954.~\cite{asakura1954interaction} The strength and range of this depletion interaction can be independently tuned by varying the polymer fugacity and  polymer size.~\cite{gast1983polymer,meijer1991computer,lekkerkerker1992phase,ilett1995phase,dijkstra1999phase}  The
possibility of tailoring the effective interactions enriches the physics of colloidal systems compared to
simple atomic fluids, and leads to a wide range of potential  applications. 

 \begin{figure}[t]
\includegraphics[scale=0.25]{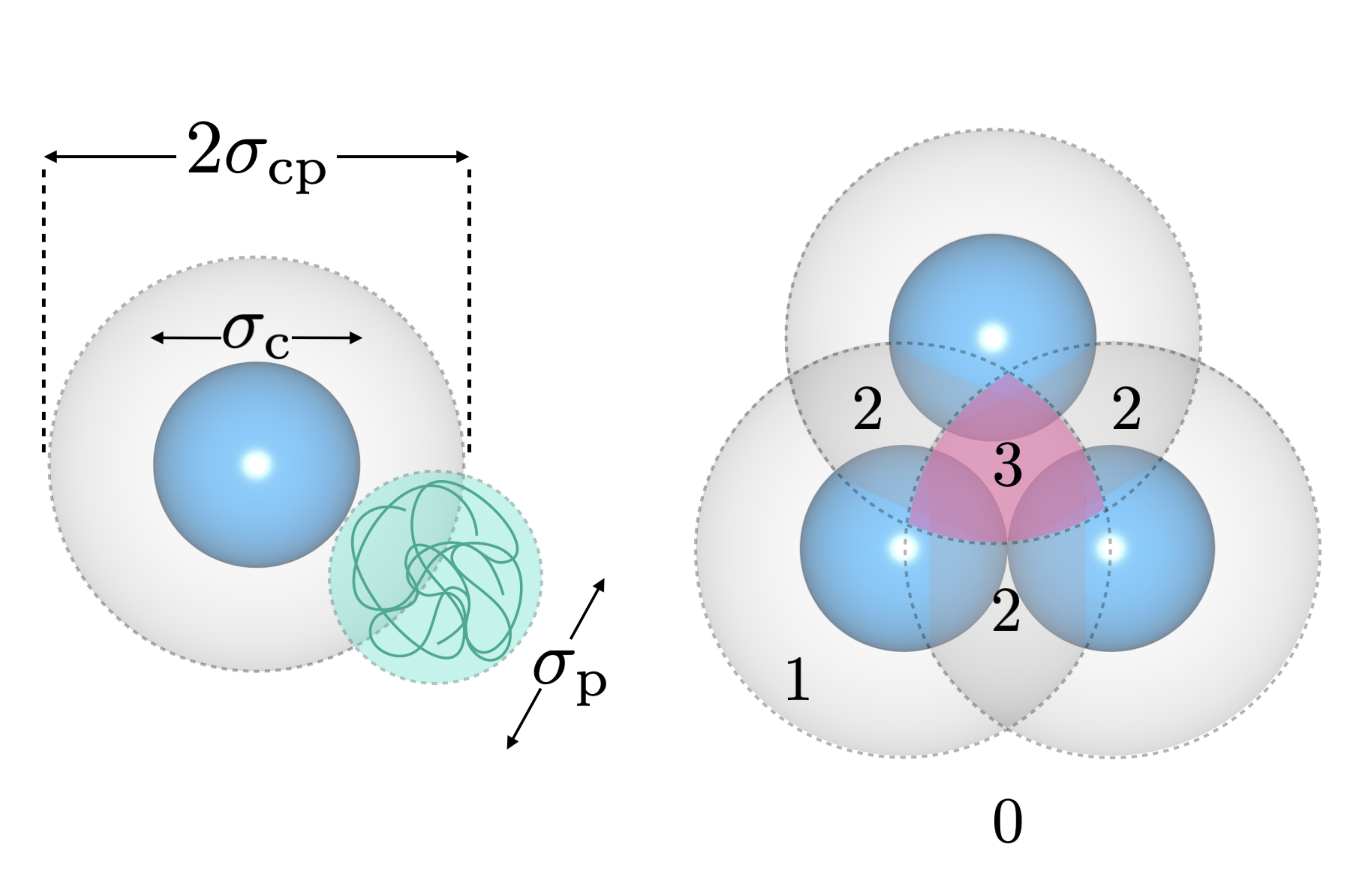}
\caption{\label{fig:pic} Left: Schematic configuration of a colloid (blue sphere) with its corresponding depletion zone (gray-shaded sphere). Right: The values of the descriptor $n=n(\boldsymbol{r})$, indicating the number of simultaneously overlapping depletion layers, are drawn for different regions in the triplet configuration. The region highlighted in pink, with $n=3$, corresponds to the  three-body correction, $V_{\text{f}}^{(3+)}$, that is added on top of the pair-wise approximation of the total free volume. In general, non-zero contributions to $V_{\text{f}}^{(3+)}$ will stem from regions where $n \geq 3$.}
\end{figure}

Computer simulations of these systems remain computationally expensive due to slow dynamics as very different length and time scales
are involved for the various species and as the number of microscopic species outweighs by orders of magnitude the number of colloidal particles.~\cite{dijkstra2001computer} Huge efforts have been devoted to speeding up equilibration in highly asymmetric mixtures by the implementation of (rejection-free) cluster moves,~\cite{dijkstra1997entropy,buhot1998numerical,lobaskin1999simulation,liu2004rejection,vink2004grand,bernard2009event} lattice discretization methods,~\cite{meijer1991computer,dijkstra1994evidence,panagiotopoulos1999large} or by exploiting a  sophisticated field-theoretic description  of the smaller species  in the external field of a fixed  colloid configuration within a density functional framework~\cite{lowen1992ab,fushiki1992molecular}  in the same spirit as the "ab initio" method of Car and
   Parrinello for ion-electron systems.~\cite{car1985unified}  However, all these approaches have their own limitations, and as a consequence higher densities, larger system sizes and larger size asymmetries are still unattainable, thereby leaving many intriguing experimental observations unexplained such as void formation, gas-liquid and gas-solid coexistence in like-charged colloidal suspensions,~\cite{tata1992vapor,ito1994void,kepler1994attractive,tata1997amorphous,larsen1997like} as well as hampering investigations of interesting phenomena like capillary wave fluctuations,  long wavelength fluctuations of marginal colloidal liquids at their triple point,  and density oscillations at the gas-liquid interface in colloid-polymer mixtures.~\cite{aarts2004direct,vink2005capillary,moussaid1999structure,vis2020quantification} 

 \begin{figure*}[t]
\includegraphics[scale=0.14]{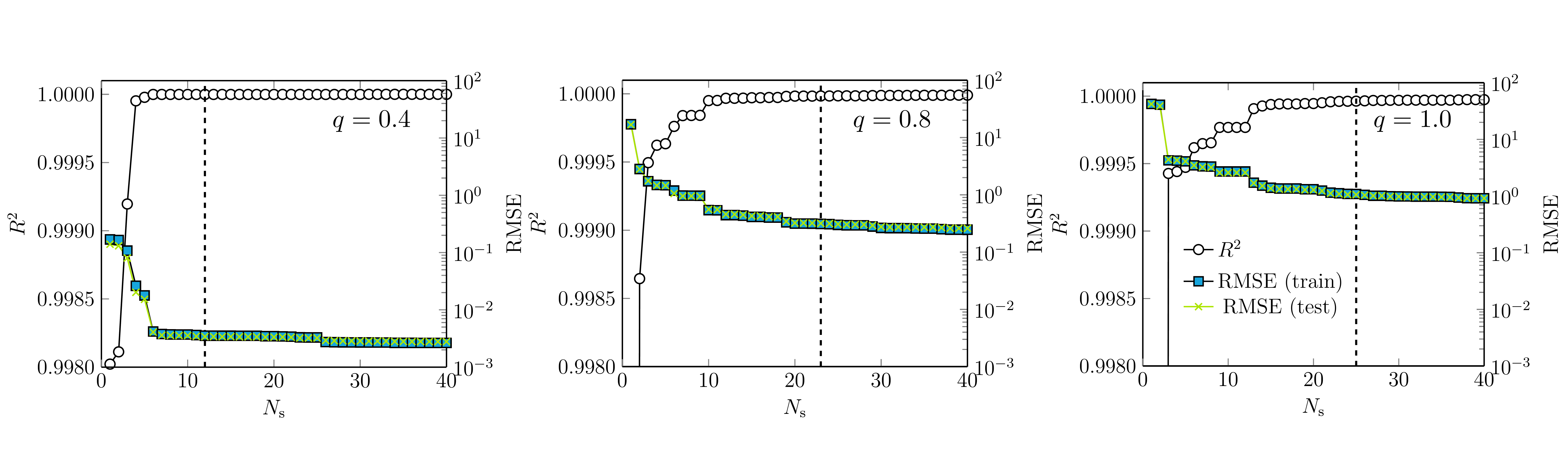}
\caption{\label{fig:qual} Square of the correlation coefficient $R^{2}$ and root mean squared error RMSE as a function of the number  of selected SFs $N_{\text{s}}$ for a colloid-polymer mixture with size ratio $q=0.4$, $0.8$ and $1.0$. The RMSE is shown for both the training and test sets. The optimal value of the number of SFs is indicated with the vertical dashed line for each case. Note that $R^{2}$ is plotted on linear scale and RMSE on logarithmic scale.}
\end{figure*}

   An alternative strategy to circumvent slow equilibration is to formally integrate out the degrees of freedom of the microscopic species in the partition sum and to derive an exact expression for the  effective one-component Hamiltonian of the colloids. The  effective Hamiltonian, which depends on all colloid coordinates and involves many-body interactions, can be employed in standard simulation schemes, but its evaluation becomes extremely computationally demanding when three- and higher-body interactions are important.~\cite{dijkstra2006effect,kobayashi2019correction} 
 
 In the past years, machine learning (ML) techniques have been employed  to efficiently approximate the  many-body interatomic potentials in atomistic systems by fitting large data sets from electronic structure calculations.~\cite{rupp2012fast,behler2016perspective,musil2021physics} These ML potentials can be used in molecular simulations, thereby combining the accuracy of ``first principles" calculations with  the efficiency of simple atomistic ``force-fields".~\cite{behler2016perspective} 
 
In this Article, we introduce  a general ML  approach to efficiently represent the effective many-body interactions in colloidal systems. We consider a model suspension of sterically-stabilized colloidal particles and non-adsorbing ideal polymers for which a wealth of data exist on the  phase behavior and structure, both from numerical evaluations of the exact one-component Hamiltonian or from direct simulations of the true binary mixture.\cite{dijkstra1999phase,dijkstra2006effect,verso2006critical,vink2004grand} Moreover, exemplifying our method using this model system is convenient as the importance of  two-, three-, and higher-body contributions to the effective potential can be tuned by the polymer size and colloid density, allowing for a systematic exploration of the validity of our ML approach. We fit the  effective many-body potentials as a function of all colloid coordinates with the symmetry functions (SFs) as introduced by Behler and Parrinello,~\cite{behler2007generalized} which were also recently used to fit the many-body interactions between microgel particles.~\cite{boattini2020modeling}  We find that the fitted ML potential reduces the computational cost by up to almost four orders of magnitude in comparison to a numerical evaluation, and accurately describes  the effective many-body interactions for a wide range of colloid densities  and polymer fugacities. Hence, the ML potential can be assumed to be state-independent and can straightforwardly be used in Monte Carlo (MC)  simulations to observe direct coexistence of a dilute  colloidal gas and a dense colloidal liquid phase. Using these ML potentials,  we find good agreement with the structure, phase behavior, and interfacial tension as obtained in previous studies for mixtures of different size ratio.

The remainder of this article is structured as follows. In Sec.~\ref{sec:model}, we describe the model colloid-polymer mixture and its formal mapping onto the corresponding effective one-component representation. The employed SFs and fitting procedure of the ML potentials are presented in Sec.~\ref{sec:fitting}, whereas the  results of the MC simulations using the ML potentials, demonstrating the accuracy of the model, are discussed in Sec.~\ref{sec:valida}. We conclude with a final discussion in Sec.~\ref{sec:conclusions}.

\section{\label{sec:model}Effective one-component description of the Asakura-Oosawa model}
The simplest model that captures the essence of  polymer-induced effective interactions between colloidal particles~\cite{asakura1954interaction,asakura1958interaction} was introduced by Vrij.~\cite{vrij1976polymers} In this so-called Asakura-Oosawa (AO) model, the colloidal particles are regarded as hard spheres of diameter $\sigma_{\text{c}}$, whereas the polymer coils with radius of gyration $R_{\text{g}}$  are treated as ideal point particles  as regards their mutual interactions. The colloid-polymer pair interaction is hard-sphere-like such that
their distance of closest approach is $\sigma_{\text{cp}} = (\sigma_{\text{c}} + \sigma_{\text{p}})/2$ with polymer diameter  $\sigma_{\text{p}}=2 R_{\text{g}}$. 
We consider a system consisting of $N_{\text{c}}$ colloidal hard spheres at  coordinates $\{\boldsymbol{R}_i\}$ with $i=1, \dots, N_{\text{c}}$,  and $N_{\text{p}}$ polymer coils at positions $\{\boldsymbol{r}_j\}$ with $j=1, \dots, N_{\text{p}}$ with a size ratio $q=\sigma_{\text{p}}/\sigma_{\text{c}}$ in a volume $V$ at temperature $T$. The pair interactions of the AO model read 
\begin{equation}
 \phi_{\text{cc}}(R_{ij}) =
    \begin{cases}
      \infty & \text{for } R_{ij} < \sigma_{\text{c}}\\
      0 & \text{otherwise, } 
    \end{cases}       
\end{equation}
\begin{equation}
 \phi_{\text{cp}}(|\boldsymbol{R}_{i}-\boldsymbol{r}_{j}|) =
    \begin{cases}
      \infty & \text{for } |\boldsymbol{R}_{i}-\boldsymbol{r}_{j}| < \sigma_{\text{cp}}\\
      0 & \text{otherwise, } 
    \end{cases}       
\end{equation}
\begin{equation}
     \phi_{\text{pp}}(r_{ij})= 0,
\end{equation}
with $R_{ij}=|\boldsymbol{R}_{i} - \boldsymbol{R}_{j}|$  the center-of-mass distance between colloid $i$ and $j$.   This binary mixture of colloids and polymer is described by the total interaction Hamiltonian  $ \mathcal{H} = \mathcal{H}_{\text{cc}} + \mathcal{H}_{\text{cp}} + \mathcal{H}_{\text{pp}} $ with 
\begin{equation}
    \mathcal{H}_{\text{cc}}= \sum_{i<j}^{N_{\text{c}}}\phi_{\text{cc}}(R_{ij}),
\end{equation}
\begin{equation}
\mathcal{H}_{\text{cp}}= \sum_{i=1}^{N_{\text{c}}}\sum_{j=1}^{N_{\text{p}}}\phi_{\text{cp}}(|\boldsymbol{R}_{i}-\boldsymbol{r}_{j}|),
\end{equation}
\begin{equation}
 \mathcal{H}_{\text{pp}}\equiv 0.
\end{equation}

\begin{figure*}[htb]
\includegraphics[scale=0.138]{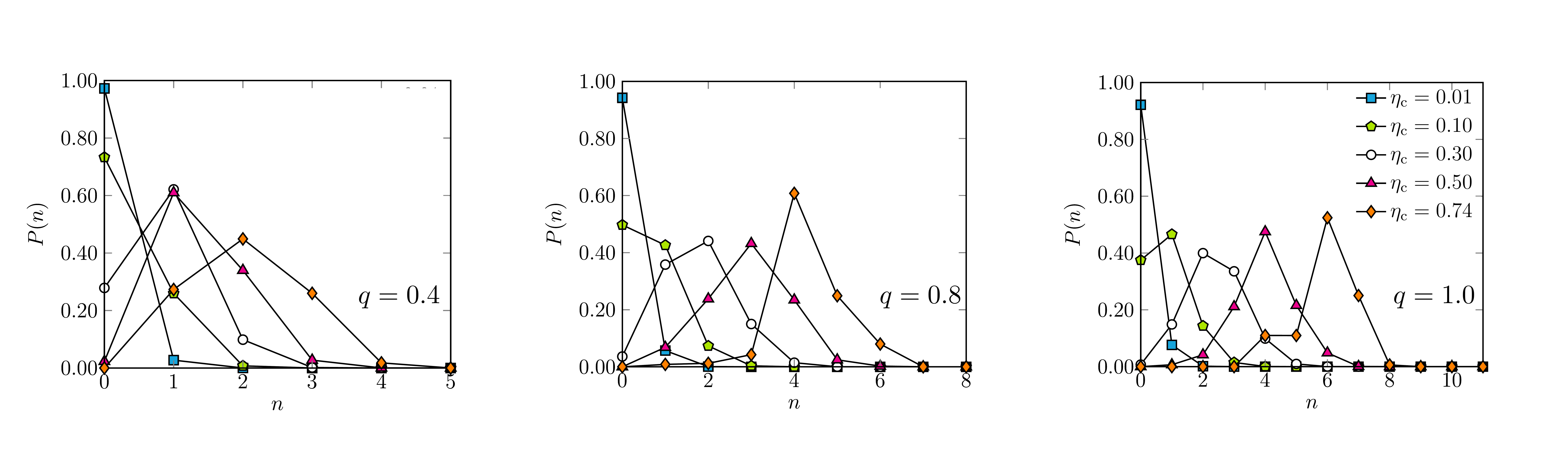}
\caption{\label{fig:distributions} The probability of $n$ overlapping depletion layers for a colloid-polymer mixture with polymer reservoir packing fraction $\eta_{\text{p}}^{\text{r}}=0$ and colloid packing fraction $\eta_{\text{c}}$ as labelled for varying size ratios $q=0.4$, $0.8$ and $1.0$.}
\end{figure*}

It is convenient to treat the polymer coils grand-canonically, in which the fugacity of the polymers $z_{\text{p}}$, or equivalently the polymer reservoir packing fraction $\eta_{\text{p}}^{\text{r}}\equiv \pi \sigma_{\text{p}}^{3} z_{\text{p}}/6$, is fixed. 
The thermodynamic potential $F(N_{\text{c}}, z_{\text{p}}, V, T)$ of this binary system reads 
\begin{widetext}
\begin{eqnarray}
\label{f}
\exp\lbrack -\beta F \rbrack &=& \sum_{N_{\text{p}}=0}^{\infty} \frac{z_{\text{p}}^{N_{\text{p}}}}{N_{\text{c}}!\Lambda_{\text{c}}^{3N_{\text{c}}}N_{\text{p}}!}\int_V d\boldsymbol{R}^{N_{\text{c}}}\int_V d\boldsymbol{r}^{N_{\text{p}}} \exp\lbrack -\beta(\mathcal{H}_{\text{cc}}+\mathcal{H}_{\text{cp}})\rbrack,  \nonumber \\
&=& \frac{1}{N_{\text{c}}! \Lambda_{\text{c}}^{3N_{\text{c}}}}  \int_{V}d\boldsymbol{R}^{N_{\text{c}}}  \exp \left[ -\beta \mathcal{H}_{\text{cc}}  \right] \exp \left[ -\beta \Omega \right] \nonumber \\
& = &  \frac{1}{N_{\text{c}}!\Lambda_{\text{c}}^{3N_{\text{c}}}}\int_V d\boldsymbol{R}^{N_{\text{c}}} \exp\lbrack -\beta \mathcal{H}_{\text{eff}}\rbrack, 
\end{eqnarray}
\end{widetext}
where $\beta = (k_{\text{B}}T)^{-1}$ with $k_{\text{B}}$ the Boltzmann constant, and  $\Lambda_{\alpha}$ is the thermal wavelength of species $\alpha=\text{c},\text{p}$. By integrating out the degrees of freedom of the polymers, this binary mixture can be mapped onto an effective one-component system described by an effective colloids-only Hamiltonian $\mathcal{H}_{\text{eff}}= \mathcal{H}_{\text{cc}}+ \Omega$, where  $\Omega = -\beta^{-1} \ln \left[  \sum_{N_{\text{p}}=0}^{\infty}\frac{z_{\text{p}}^{N_{\text{p}}}}{N_{\text{p}}!}   \int_{V}d\boldsymbol{r}^{N_{\text{p}}}    \exp \left(-\beta \mathcal{H}_{\text{cp}}  \right) \right]$ denotes the grand potential of a ``sea" of ideal polymers at fugacity $z_{\text{p}}$ in the external field of a fixed configuration of $N_{\text{c}}$ colloids.~\cite{dijkstra1999phase,dijkstra2006effect,dijkstra2002entropic} For the AO model, the grand potential $\Omega$ is simply the negative of the  free volume available for the polymer in the fixed configuration of $N_{\text{c}}$ colloids, i.e. \begin{equation}
\label{om_eq}
\Omega = -z_{\text{p}} V_{\text{f}}( \{\boldsymbol{R}_i\}),
\end{equation}
which is the volume outside the $N_{\text{c}}$ depletion zones, and which can be decomposed into a zero-, one-, two-, three-, and higher-body contribution
\begin{equation}
    V_{\text{f}}= V_{\text{f}}^{(0)} + \sum_{i=1}^{N_{\text{c}}}{V_{\text{f}}^{(1)}(\boldsymbol{R}_{i})} + \sum_{i<j}^{N_{\text{c}}}{V_{\text{f}}^{(2)}(\boldsymbol{R}_{i}, \boldsymbol{R}_{j})} + V_{\text{f}}^{(3+)}, 
\end{equation}
i.e. the number   of colloids $k=0, 1, 2,\cdots, N_{\text{c}}$ that interact simultaneously with the ``sea'' of ideal polymer, and where  $V_{\text{f}}^{(3+)}$ denotes the three- and higher-body term. Analytical expressions exist for $k=0,1,$ and 2. The zero-body contribution $V_{\text{f}}^{(0)}=V$ simply corresponds to the  volume of the system, $V_{\text{f}}^{(1)}=-v_1$ is the volume excluded to a polymer by a single colloid with $v_1=\pi \sigma_{\text{cp}}^{3}/6$,  and $V_{\text{f}}^{(2)}= 8v_1(1-3x/4+x^3/16)$ with $x=R_{ij}/\sigma_\text{cp}<2$ is the well-known depletion potential of the AO model, representing the lens-shaped overlap volume of two colloidal spheres of diameter $\sigma_{\text{cp}}$ at separation $R_{ij}$. 

For $q <  0.1547$, the higher-order contributions incorporated in $V_{\text{f}}^{(3+)}$ are  zero, and hence a  mapping onto an effective one-component system with an effective Hamiltonian based on pairwise additive depletion potentials is exact.~\cite{gast1983polymer,dijkstra1999phase} For larger $q$, the model based solely on the pairwise approximation exhibits a phase behaviour that strongly differs from that observed in the system where three- and higher-body interactions $V_{\text{f}}^{(3+)}$ are considered. The main effect of the many-body interactions in colloid-polymer mixtures is to enhance the  $z_{\text{p}}$ regime of stable gas-liquid coexistence  at the expense of that of the gas-solid coexistence.~\cite{dijkstra1999phase,dijkstra2006effect} $V_{\text{f}}^{(3+)}$ can be computed by measuring the overlap volume of three or more depletion zones  which can only be evaluated numerically as shown in Ref. \onlinecite{dijkstra2006effect}. The key ingredient in the numerical  evaluation of $V_{\text{f}}^{(3+)}$ in Ref. \onlinecite{dijkstra2006effect} is the introduction of a spatial descriptor $n=n(\boldsymbol{r})$, which for a given colloid configuration, measures the number of simultaneously overlapping depletion layers at spatial coordinate $\boldsymbol{r}$. In Fig.~\ref{fig:pic} we show a pictorial representation of a configuration of colloidal particles (solid blue spheres) and their depletion zones (surrounding gray-shaded spheres). The values of $n$ in different regions of space are also reported. Upon the introduction of $n(\boldsymbol{r})$, the many-body correction $V_{\text{f}}^{(3+)}$ can be computed as
\begin{equation}
\label{vf3}
    V_{\text{f}}^{(3+)}= -\frac{1}{2}\int_{n\geq 3} {d\boldsymbol{r} \left[n(\boldsymbol{r}) - 1 \right]\left[n(\boldsymbol{r}) - 2 \right]},
\end{equation}
where the integration is performed over the regions with $n\geq 3$.~\cite{dijkstra2006effect} Conceptually, $V_{\text{f}}^{(3+)}$ can be seen as a correction to the overestimated overlap volume of the depletion zones in $V_{\text{f}}^{(2)}$. 

\section{\label{sec:fitting}Fitting the many-body potential}
As described above, $V_{\text{f}}^{(3+)}$ is a function of the colloid configuration $\boldsymbol{R}^{N_{\text{c}}}$. Therefore, its evaluation requires to sample the whole volume $V$ in order to identify those regions where $n \geq 3$. In practice, $n(\boldsymbol{r})$ can be measured using a spherically-symmetric grid of $M$ points around each colloid. The number of  points needs to be sufficiently high (on the order of $M \sim 10^{5}$) to accurately sample $V_{\text{f}}^{(3+)}$, making its implementation computationally feasible only for a few hundreds of colloids. Here, we use an ML approach to fit  $V_{\text{f}}^{(3+)}$ as a function of all colloid coordinates using a set of SFs introduced by Behler and Parrinello.~\cite{behler2011atom,behler2007generalized}

\subsection{\label{sec:dataset}Training data set}

Since $V_{\text{f}}^{(3+)}$ depends on size ratio $q$, we use a different ML fit for each  $q$. To build the training data sets, we perform MC simulations on $N_{\text{c}}=108$ colloids for $q=0.4, 0.8$, and 1.0,  polymer fugacity $z_{\text{p}}=0$, and colloid packing fraction $\eta_{\text{c}}=\pi  \sigma_{\text{c}}^3 N_{\text{c}}/6V \in \left[ 0.15, 0.65\right]$ with a packing fraction spacing of $\delta \eta_{\text{c}}=0.005$.  From each simulation we collect 500 equilibrated, essentially uncorrelated configurations\footnote{The configurations we use are spaced far enough apart in number of MC cycles that the particles have had a chance to diffuse significantly between successive snapshots. Even in the high density fluid phase at $\eta_{\text{c}}=0.40$, the particles travel more than 6 times their diameters between successive configurations.} and measure $V_{\text{f}}^{(3+)}$ using a spherically-symmetric $(r^{3}, \cos (\theta), \phi)$ grid of 100, 50 and 50 points, respectively.~\cite{dijkstra2006effect} The resulting data set for each $q$ contains a total of $49,500$ representative  particle configurations at different colloid densities, from which $80\%$ are used for training and $20\%$ for testing. It is worth mentioning that the selected case of zero polymer fugacity corresponds to a single-component system of hard spheres, which is computationally extremely inexpensive to simulate. Therefore, equilibration in this system is achieved rapidly and we were able to access a large number of decorrelated configurations with relevant local environments of the individual particles in low- and high-density fluid and solid phases, which are also present in colloid-polymer mixtures. Learning $V_{\text{f}}^{(3+)}$ instead of the full grand potential $\Omega$ significantly reduces the size of the training data set and lowers the computational effort while still giving access to non-zero polymer fugacity via Eq.~\ref{om_eq}.

\begin{figure}
\includegraphics[scale=0.68]{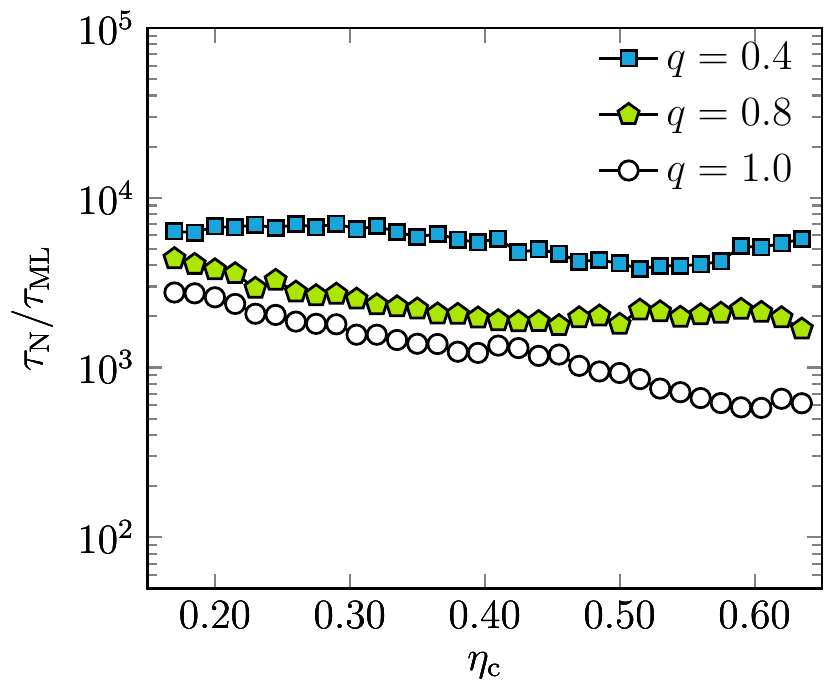}
\caption{\label{fig:speed} Ratio between the computational time required for  the numerical evaluation of the many-body term term  ($\tau_\text{N}$) and the ML  potential ($\tau_\text{ML}$) as a function of colloid packing fraction $\eta_{\text{c}}$ of a colloid-polymer mixture with varying size ratio $q$ and polymer fugacity $z_{\text{p}}=0$.}
\label{fig:speedup}
\end{figure}

\begin{figure*}
\includegraphics[scale=0.48]{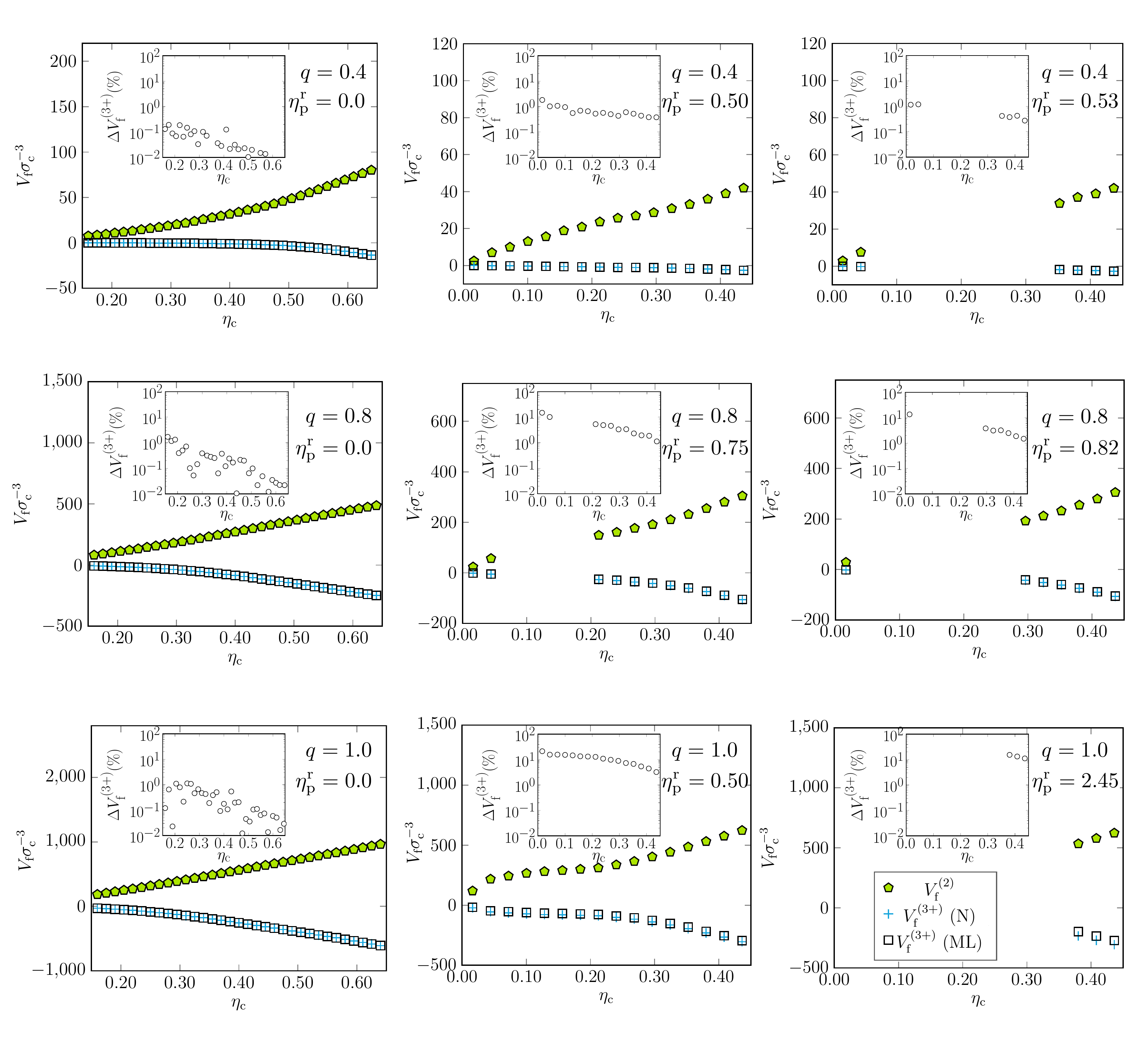}
\caption{\label{fig:vf} The two-body $V_{\text{f}}^{(2)}$ (pentagons), and three- and higher-body $V_{\text{f}}^{(3+)}$ contribution to the effective many-body potential as predicted by the ML potential (ML) (squares) for a colloid-polymer mixture with size ratio $q=0.4, 0.8$, and 1.0 as a function of colloid packing fraction $\eta_{\text{c}}$ and varying polymer reservoir packing fraction $\eta_{\text{p}}^{\text{r}}$ along with the actual $V_{\text{f}}^{(3+)}$ as obtained from a numerical evaluation (N) (plus). Inset plots correspond to the percentage errors of the many-body term, $\Delta V_{\text{f}}^{(3+)}= \left[|V_{\text{f}}^{(3+)}(\text{N}) - V_{\text{f}}^{(3+)}(\text{ML}) | \right]/ V_{\text{f}}^{(3+)}(\text{N}) \times 100 $. Note that for $\eta_{\text{p}}^{\text{r}} \neq 0 $ we consider only state points outside coexistence regions.}
\end{figure*}

\begin{figure*}
\includegraphics[scale=0.485]{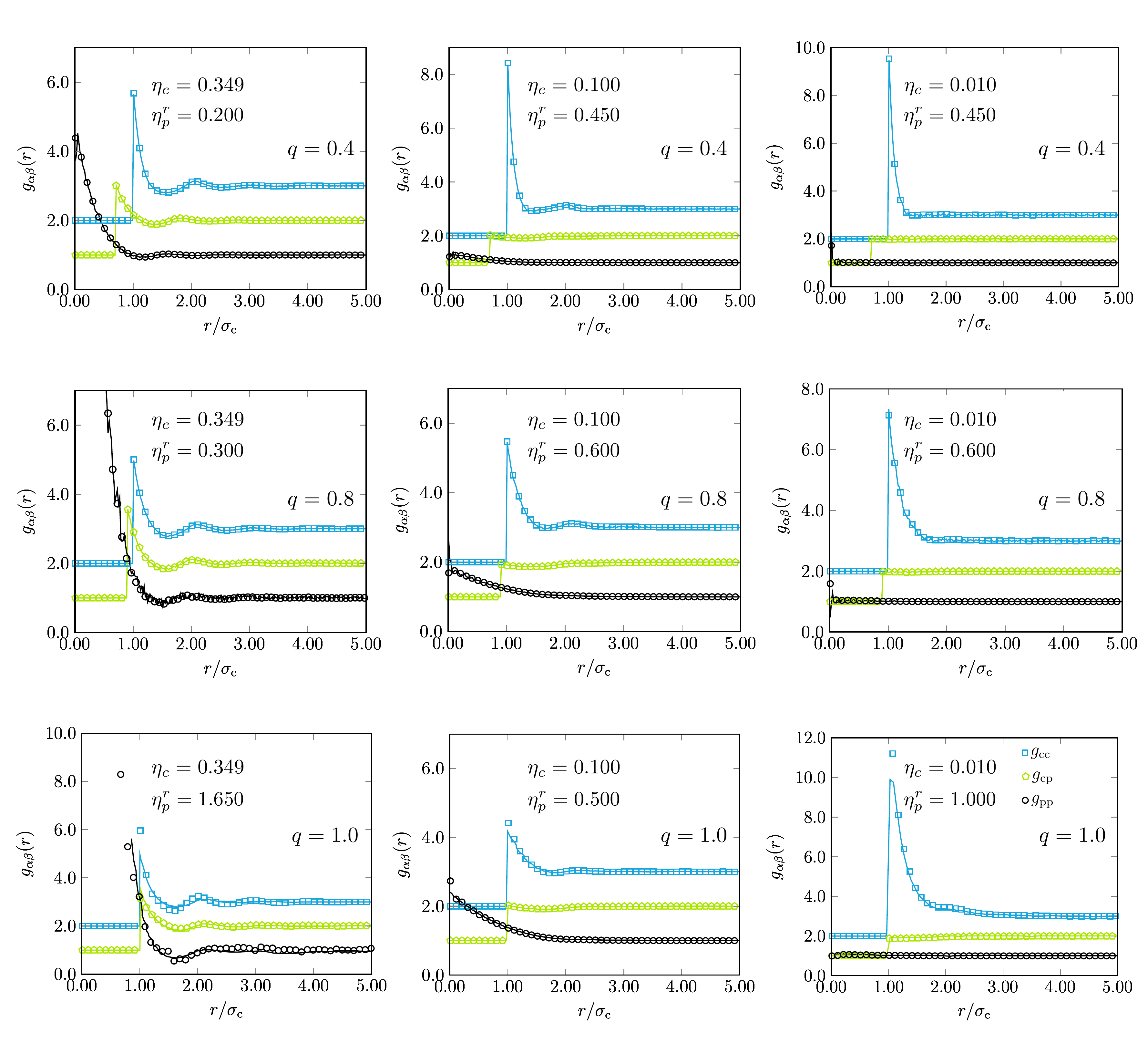}
\caption{\label{fig:grs} Colloid-colloid (squares), colloid-polymer (pentagons) and polymer-polymer (circles) pair correlation functions $g_{\alpha \beta}(r)$ with $\alpha,\beta=\text{c},\text{p}$ of a colloid-polymer mixture with $q=0.4$, $0.8$ and $1.0$ obtained from MC simulations using the ML potential at $\eta_{\text{c}}$ and $\eta_{\text{p}}^{\text{r}}$ as labeled. In the top and middle panels, the solid lines are the results obtained from MC simulations based on a numerical evaluation of $V_{\text{f}}^{(3+)}$, whereas in the bottom panel, the solid lines correspond to the results reported in Ref. \onlinecite{dijkstra2006effect}. For clarity, $g_{\text{cp}}(r)$ and $g_{\text{cc}}(r)$ are shifted in the vertical direction. }
\end{figure*}

\subsection{\label{sec:symfs}Symmetry functions}
To describe the local environment of a particle we use the SFs introduced by Behler and Parrinello for constructing high-dimensional neural network potentials.~\cite{behler2007generalized} These SFs are described in great detail in Refs. \onlinecite{behler2011atom,behler2015constructing} and have been used as inputs for atomic feed-forward neural networks in order to provide the atomic energy contributions of different materials and molecules.~\cite{behler2007generalized,khaliullin2010graphite,eshet2012microscopic,kapil2016high,cheng2016nuclear} Recently, they have also been used in combination with linear regression to fit effective many-body interactions between elastic spheres.~\cite{boattini2020modeling} 

Since the three- and higher-body term $V_{\text{f}}^{(3+)}$ in Eq. \ref{vf3} does not include any two-body contributions, we consider only the angular three-body SFs $G^3(i)$ for particle $i$, which are defined as 

\begin{eqnarray}
    G^{3}(i) =&& 2^{1-\xi}\sum_{j,k\neq i}\left( 1 + \lambda \cos \theta_{ijk} \right)^\xi e^{-\eta\left( R_{ij}^{2} + R_{ik}^{2} +R_{jk}^{2} \right)}\nonumber\\
    &&\times f_{c}(R_{ij})f_{c}(R_{ik})f_{c}(R_{jk}),
\label{eq:g3}
\end{eqnarray}
where the indices $j$ and $k$ run over all the neighbours of particle $i$, and $\xi$, $\eta$, and $\lambda$ are three parameters that determine the shape of the function. The parameter $\lambda$ can have the values $+1$ or $-1$ and determines the angle $\theta_{ijk}$ at which the angular part of the function has its maximum. The angular resolution is provided by the parameter $\xi$, while $\eta$ controls the radial resolution. Additionally, $f_c(R_{ij})$ is a cutoff function: a monotonically decreasing function that smoothly goes to 0 in both value and slope at the cutoff distance $r_c$. Here, we consider a cutoff function of the form
\begin{equation}
 f_{c}(R_{ij}) =
    \begin{cases}
      \tanh^{3} (1 - R_{ij}/r_{c}) & \text{for } R_{ij} \leq r_{c}\\
      0 & \text{for } R_{ij}>r_{c}.
    \end{cases}       
\end{equation}
Using the training set consisting of representative colloid configurations at different colloid densities, we fit $V_{\text{f}}^{(3+)}$ with a linear combination of $N_{\text{s}}$ SFs in Eq. \ref{eq:g3}, and select the optimal subset of SFs using the feature selection scheme of Ref.  \onlinecite{boattini2020modeling}, which we summarize in the following. We note that we also implemented other feature selection schemes such as those introduced by Imbalzano et al.,~\cite{imbalzano2018} but the one of Ref.  \onlinecite{boattini2020modeling} turns out to be more efficient (see Appendix).

The first step of the method involves the creation of a large but manageable pool of candidate SFs. This is done by calculating, for every configuration in the data set, several SFs with different sets of parameters. Specifically, we generate the $G^{3}(i)$ SFs by setting  $r_c=2\sigma_{\text{cp}}$, $\lambda\in\{-1,1\}$, $\eta\in \{0.001,0.01,0.1,1,2,4,8\}$, and $\xi\in \{1,2,4,8\}$. With these choices, our pool of candidates consists of $D=56$ SFs.

Then, an optimal subset of $N_{\text{s}}<D$ SFs is selected from the pool in a step-wise fashion. First, the  SF is selected with the largest correlation with the many-body term as quantified by the Pearson correlation coefficient $c_{k}$

\begin{equation}
    c_{k}=\frac{ \sum_{j}{\left (\sum_i G_k^{3}(i)|_{j} -\overline{  \sum_i G_k^3(i)}\right)\left(V_{\text{f}}^{(3+)}|_{j}-\overline{ V_{\text{f}}^{(3+)}}\right)}}{\sigma_{\text{SD}}(\sum_i G_k^{3}(i)) \sigma_{\text{SD}}(V_{\text{f}}^{(3+)})},
\label{eq:pearson}
\end{equation}
where $\sum_i G_k^{3}(i)|_{j}$ represents the sum of the $k$-th SF over all colloidal particles $i$ in configuration $j$ and $V_{\text{f}}^{(3+)}|_{j}$ denotes the many-body correction evaluated for this configuration. $\overline{  \sum_i G_k^3(i)}$ and $\overline{ V_{\text{f}}^{(3+)}}$ correspond to arithmetic means over the whole data set, and $\sigma_{\text{SD}}(\sum_i G_k^{3}(i))$ and $\sigma_{\text{SD}}(V_{\text{f}}^{(3+)})$ to their standard deviations. The next SF is then selected based on the highest increase  in the linear correlation between the currently selected set and the target many-body term as determined by 
\begin{equation} 
R^2 = {\bf c}^T {\bf R}^{-1} {\bf c},
\end{equation}
where ${\bf c}^T = (c_1, c_2,\cdots)$ is the vector whose $j$-th component is given by the Pearson correlation coefficient (\ref{eq:pearson}) between the $j$-th SF and the many-body term, and ${\bf R}$ is the correlation matrix of the current set of SFs with elements $\mathcal{R}_{ij}$ representing the Pearson correlation function between the $i$-th and $j$-th SF.  
This choice guarantees that only SFs that add relevant information are selected, while penalizing  highly correlated SFs with only redundant information as well as SFs that are sensitive to aspects of the particle’s environment that poorly correlate with the target many-body term. This process is repeated iteratively and new SFs are selected until the correlation stops increasing appreciably. Finally, the selected subset of SFs is used to approximate the target many-body term via simple linear regression.

\subsection{\label{sec:Val}Accuracy of the ML potentials}
In Fig.~\ref{fig:qual} we report the correlation coefficient $R^{2}$ and the root mean squared error
(RMSE) of the linear fits with the actual $V_{\text{f}}^{(3+)}$ as a function of the number  of SFs  for $q=0.4, 0.8,$ and 1.0, for both the training and the test set. We note that RMSE and $R^2$ are related by  a simple relation  $R^2 = 1-\text{RMSE}^2/\sigma^2_{\text{SD}}(V_{\text{f}}^{(3+)})$. Upon increasing $q$, we clearly observe that an increasing  number of SFs is required to accurately approximate $V_{\text{f}}^{(3+)}$, which can  be understood as the thickness of the depletion layers increases with $q$, thereby  enhancing the many-body effects. To quantify the importance of the many-body contributions to the effective potential, we calculate $P(n)$, the probability   that we find  $n=n(\boldsymbol{r})$ overlapping depletion layers at spatial coordinate $\boldsymbol{r}$ in a system  of $N_{\text{c}}$ colloids in a volume $V$ at 
polymer fugacity $z_{\text{p}}$, and size ratio $q$. In Fig.~\ref{fig:distributions} we show $P(n)$ for the three considered size ratios $q=0.4, 0.8$, and 1.0 at  varying packing fraction $\eta_{\text{c}}$ and $\eta_{\text{p}}^{\text{r}}=0$. For $q=0.4$, we find that the largest number $n$ with non-zero probability is $n=4$, and hence the effective potential consists of up to four-body contributions. For $q=0.8$ and $1.0$, we find that 6- and 7-body contributions  become  non-negligible at high colloid densities. We choose $N_{\text{s}}=12, 23,$ and 25 for $q=0.4, 0.8,$ and 1.0, respectively, for which we find good agreement between the ML fits and the actual many-body $V_{\text{f}}^{(3+)}$ term.

\subsection{\label{sec:eff}Efficiency of the ML potentials}
To quantify the efficiency of our ML potentials, we determine the ratio between the computational time for the numerical evaluation of the many-body term   ($\tau_\text{N}$) and the ML potential ($\tau_\text{ML}$) for different values $\eta_{\text{c}}$ and $q$. The ratios $\tau_\text{N}/\tau_\text{ML}$ are extracted by selecting 10 decorrelated configurations for each colloid packing fraction $\eta_{\text{c}}$ from the training data set, evaluating the serial computing times for the calculation of the many-body term and repeating this procedure 100 times to get the average ratios. Note that the implementations of both algorithms was simple and no neighbour lists were used. Moreover, both codes were compiled with the same compiler optimizations. From Fig. 4, we  find that our ML potentials speed up the many-body term evaluation at least by two  orders of magnitude for size ratio $q=1.0$  up to almost four orders of magnitude for $q=0.4$. The speed up decreases with colloid density $\eta_{\text{c}}$ (due to the increasing number of neighbours) and with polymer size as the cut-off value of the interaction depends on $q$. The observed speed up achieved with our ML potentials can be rationalized by comparing the order of the computations on which both algorithms are based. When the many-body correction is evaluated in a system of $N_{\text{c}}$ colloids through numerical integration, the time needed for such a computation scales as $N_{\text{c}}^{2}M$, with $M$ the number of grid points. In contrast, if no tricks are used, the time required for the evaluation of $V_{\text{f}}^{(3+)}$ using the three-body SFs scales as $N_{\text{c}}^{3}$. Thus, as it is verified in our case, a significant speed up using the ML potentials can be achieved as long as $N_{\text{c}}< M$.

\begin{figure}
\includegraphics[scale=0.24]{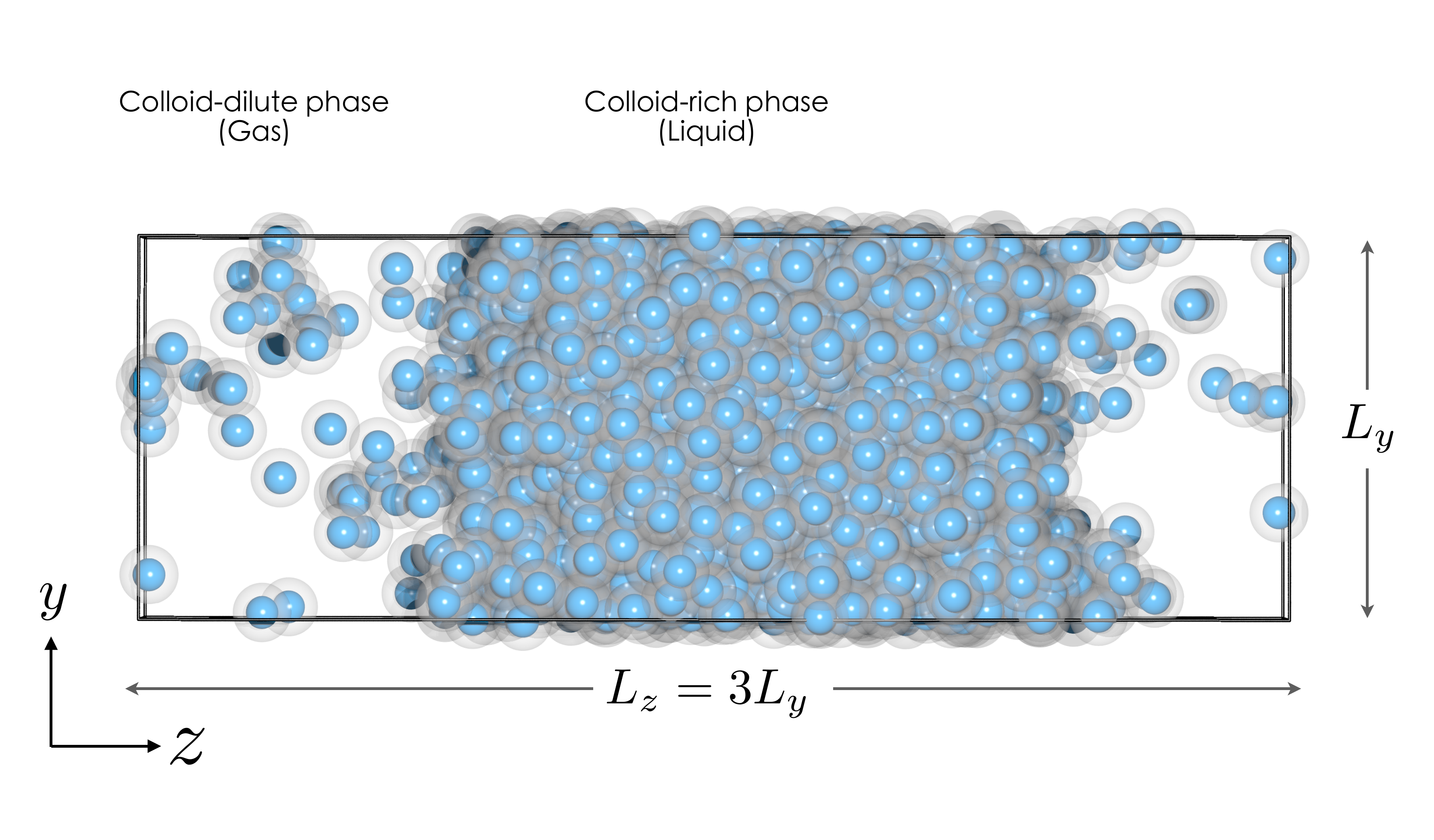}
\caption{\label{system} Typical  configuration of a colloidal gas-liquid phase coexistence of a colloid-polymer mixture as obtained from direct coexistence simulations using the ML potential.  The blue solid spheres represent the colloidal particles, whereas the gray layers surrounding the colloids illustrate their depletion zones. The polymers are integrated out and invisible.}
\end{figure}

\begin{figure*}
\includegraphics[scale=0.49]{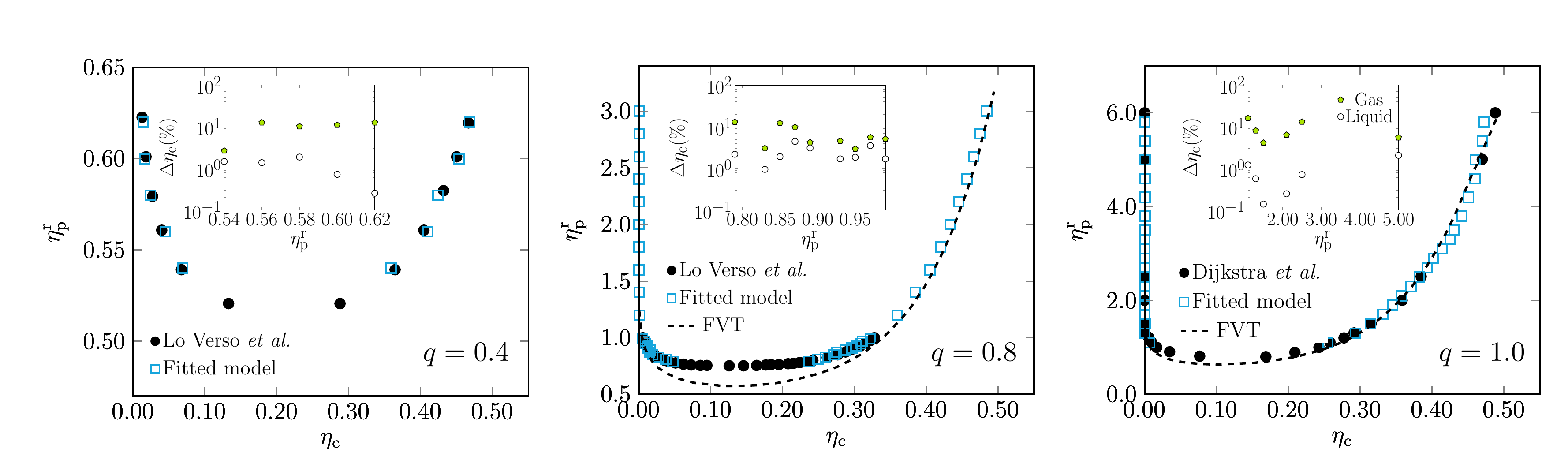}
\caption{\label{fig:vleq} Colloidal gas-liquid binodals in the colloid packing fraction $\eta_\text{c}$ - polymer reservoir packing fraction $\eta_{\text{p}}^{\text{r}}$ plane of a colloid-polymer mixture with size ratio $q=0.4$, $0.8$ and $1.0$ as obtained from  direct-coexistence simulations using the ML potentials (open squares)  and from earlier reports (solid circles),  either by Dijkstra et al.~\cite{dijkstra2006effect} using a numerical evaluation of the effective many-body interactions or by Lo Verso et al.~\cite{verso2006critical} employing simulations of the true binary mixture. The dashed lines represent the binodals predicted by free-volume theory (FVT).\cite{dijkstra1999phase,dijkstra2006effect} Inset plots correspond to the percentage error of the average density of the coexisting gas (solid pentagons) and liquid (empty circles) phases at fixed $\eta_{\text{p}}^{\text{r}}$, $\Delta \eta_{\text{c}}= \left[|\eta_{\text{c}}(\text{N}) - \eta_{\text{c}}(\text{ML}) | \right]/ \eta_{\text{c}}(\text{N}) \times 100 $.}
\end{figure*}

\section{\label{sec:valida}Validation}

In the following, we focus on the validation of the effective one-component ML many-body potential. It is important to note that we fit the $V_{\text{f}}^{(3+)}$ term  for a fixed $q$ with a single ML potential for all colloid packing fractions $\eta_{\text{c}} \in \left[ 0.15, 0.65\right]$ and $\eta_{\text{p}}^{\text{r}}=0$. We now investigate  the transferability of the ML potentials to state points outside the training set by considering colloid configurations at finite non-zero polymer fugacities $z_{\text{p}}$. In Fig. \ref{fig:vf}, we show the ensemble average of $V_{\text{f}}^{(3+)}$ of a colloid-polymer mixture with size ratio $q=0.4, 0.8$, and 1.0 as a function of colloid packing fraction $\eta_{\text{c}}$ and varying polymer reservoir packing fraction $\eta_{\text{p}}^{\text{r}}$ as obtained from independent MC simulations of $N_{\text{c}}=108$ colloidal particles using the ML potential along with the actual $V_{\text{f}}^{(3+)}$ as obtained from a numerical evaluation. The values of $V_{\text{f}}^{(2)}$ are also included to appreciate the importance (magnitude) of the many-body and two-body terms at different state points. We find good agreement between the ML predictions and the actual many-body $V_{\text{f}}^{(3+)}$ term for the whole range of colloid packing fractions $\eta_{\text{c}}$ and  polymer fugacities $z_{\text{p}}$, demonstrating the transferability to finite $z_{\text{p}}$ outside the training set. Some deviations are observed for $q=1.0$ at high $\eta_{\text{c}}$, where four- and higher-body interactions are  predominant.  Hence, the ML potential  can be assumed to be effectively state-independent, allowing us to simulate direct coexistence of an extremely dilute colloidal gas phase  with a very dense colloidal liquid phase in a single simulation box, see Fig. \ref{system}.


\begin{figure} 
\includegraphics[scale=0.18]{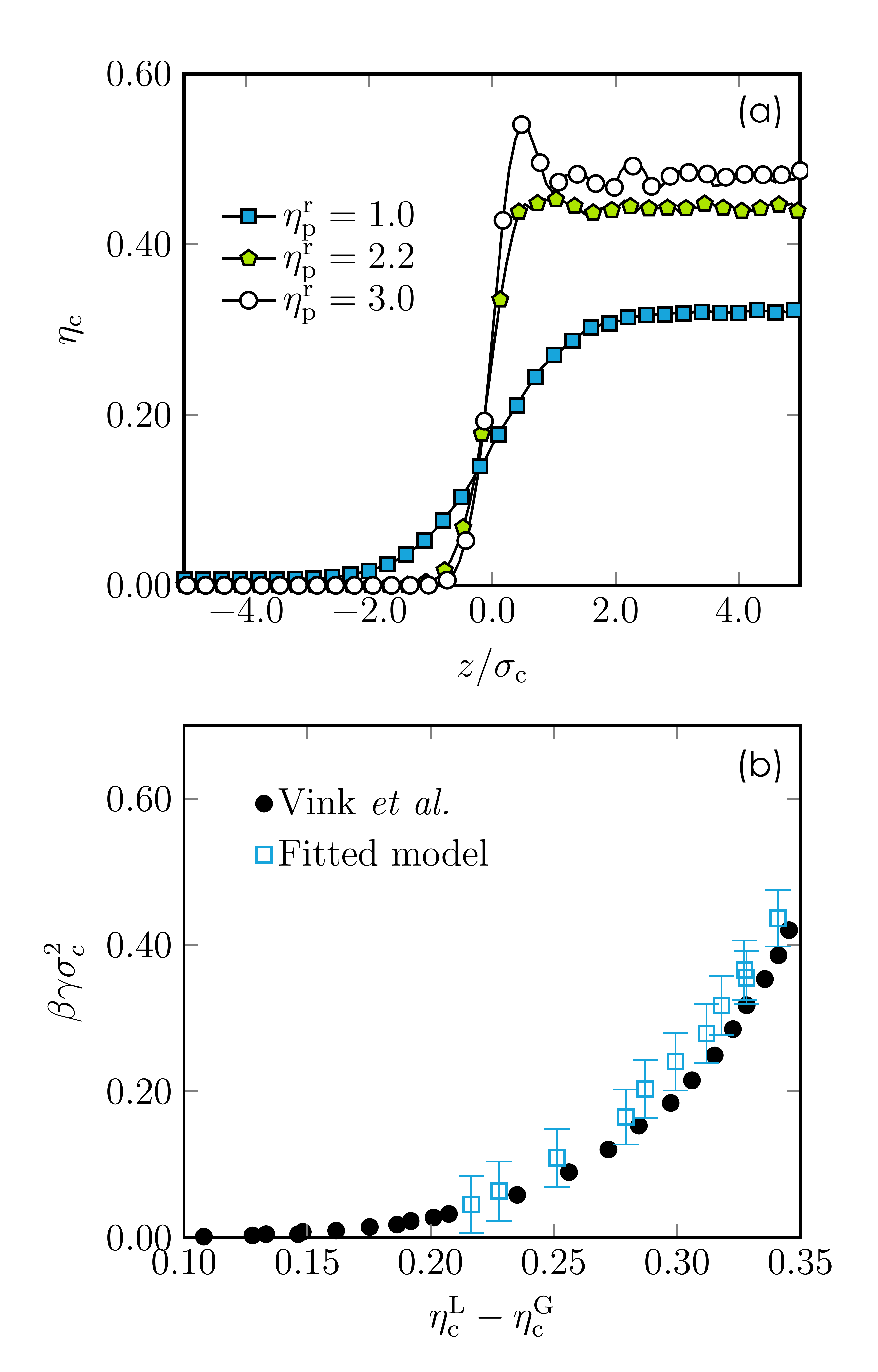}
\caption{\label{gamma08} Colloid packing fraction profiles for the free interface between coexisting gas and liquid phases of a colloid-polymer mixture with size ratio $q=0.8$ at different polymer reservoir packing fraction $\eta_{\text{p}}^{\text{r}}$, showing weak oscillations at the "colloid-rich" liquid side near the triple point (a). Gas-liquid interfacial tension $\beta \gamma \sigma_{\text{c}}^{2}$ for the same mixture as a function of  the difference in gas and
liquid packing fractions $\eta_{\text{c}}^{\text{L}}-\eta_{\text{c}}^{\text{G}}$ as obtained from direct-coexistence simulations using the ML potential (open squares) and reported by  Vink et al.~\cite{vink2004grand} (solid circles) (b).
}
\end{figure}

Given that the ML potentials accurately describe the many-body $V_{\text{f}}^{(3+)}$ term at different state points, we now test the ability of the ML potential in reproducing the structure of a colloid-polymer mixture. To this end, we measure the colloid-colloid, polymer-polymer and colloid-polymer pair-correlation functions $g_{\alpha \beta}(r)$ with $\alpha,\beta=\text{c},\text{p}$ from MC simulations of $N_{\text{c}}=1372$ particles interacting  with the ML potential at three different state points for each size ratio  $q=0.4$, 0.8 and 1.0. At first sight it may seem surprising that it is possible to recover information about the structure of the polymer as we traced out the polymer degrees of freedom. However, as the polymer coils are ideal, the number density of the polymer is constant in the free volume (or holes) of the system.\cite{dijkstra2006effect} Therefore, we can determine the colloid-polymer and polymer-polymer pair correlation functions from $10,000$ randomly inserted polymer coils in an instantaneous colloid configuration, provided that no overlap exists between the polymer and colloids. 
In Fig.~\ref{fig:grs} we plot $g_{\alpha \beta}(r)$ denoted by the symbols for the three size ratios considered at varying state points along with the ones obtained by a numerical evaluation of $V_{\text{f}}^{(3+)}$ or from previous work~\cite{dijkstra2006effect} as denoted by the solid lines. The agreement between the pair correlation function as obtained using the ML potentials with the "exact" results is evident in all cases, especially for mixtures with $q=0.4$ and 0.8, which reflects the ability of our ML potential to correctly describe the structure of the mixture, regardless of $\eta_{\text{p}}^{\text{r}}$ and $\eta_{\text{c}}$. However, it is important to note that mapping of four- and higher-body potentials onto a linear combination of angular three-body SFs does not  necessarily guarantee that the structure  of the colloid-polymer mixture is well captured, see e.g. Refs. \onlinecite{dijkstra2000effective,louis2002beware}. Indeed, small deviations are observed for the contact value of the colloid-colloid pair correlation function $g_{cc}(r)$ for the highest $q=1.0$, where four- and higher-body interactions become dominant.

As the ML potentials accurately describe $V_{\text{f}}^{(3+)}$ and the structure of colloid-polymer mixtures, we now investigate whether the ML potentials can  be used to determine the coexistence densities of the colloidal gas and liquid phases using direct-coexistence simulations. To this end, we perform MC simulations  of  $N_{\text{c}}=1372$ particles in an elongated box. We determine the packing fraction of the coexisting gas and liquid phases,  $\eta_{\text{c}}^{\text{L}}$ and $\eta_{\text{c}}^{\text{G}}$, by measuring the equilibrium density profiles. We plot the resulting gas-liquid binodals in Fig. ~\ref{fig:vleq} for colloid-polymer mixtures with  size ratio $q=0.4, 0.8$, and 1.0. For  $q=0.4$  and $0.8$, we compare our results with those obtained from  grand-canonical MC simulations using the full binary mixture,~\cite{verso2006critical} whereas the results for $q=1.0$ are compared with those obtained using the full effective one-component system.~\cite{dijkstra2006effect} Our ML results show good agreement with the previously obtained gas-liquid binodals from the critical point at low $\eta_{\text{p}}^{\text{r}}$ all the way to the triple point at high $\eta_{\text{p}}^{\text{r}}$,  even for $q=1.0$ where the four- and higher-body terms contribute for more than 75\% to the effective potential at the triple point as shown in Fig.~\ref{fig:distributions}. This is an interesting result, as we expect a reduced correspondence at large $q$ as the number of depletion layers that can simultaneously overlap increases due to larger depletion layers. In addition, we plot the results from free-volume theory for $q=0.8$ and 1.0 and find good agreement, except near the critical point where the free-volume theory is less accurate.~\cite{lekkerkerker1992phase,dijkstra2006effect} In Fig. \ref{gamma08}.a, we show the $\eta_{\text{c}}$ profiles for $q=0.8$. Surprisingly, we find weak oscillations at the "colloid-rich" liquid side of the density profiles near the triple point. These oscillations were predicted in fundamental measure theory,~\cite{brader2001entropic} but were not observed in both a square-gradient density functional approach~\cite{brader2000fluid} and in recent experiments on colloid-polymer mixtures.~\cite{vis2020quantification} Here, we show for the first time such density oscillations, although weak, at a free gas-liquid interface of a colloid-polymer mixture. The amplitude of these oscillations are reduced significantly by thermally induced capillary-wave fluctuations,~\cite{aarts2004direct} making the oscillations hard to detect in experiments~\cite{vis2020quantification} and simulations.~\cite{chacon2001layering} Such capillary wave fluctuations are related to the ultra-low interfacial tensions of colloidal systems.~\cite{brader2001entropic}

To investigate if our ML potentials can also be used to measure the interfacial gas-liquid tension $\gamma$, we use the so-called Test Area MC technique (TAMC), which is based on thermodynamic perturbation theory.~\cite{gloor2005test,vega2007surface} In particular, we sample $\gamma$ in our MC simulations by performing test area perturbations once every cycle and averaging over $5\times 10^{5}$ cycles.  We report the interfacial tension $\beta \gamma \sigma_{\text{c}}^{2}$ as a function of the difference of the coexisting densities of the gas and liquid phases  $\eta_{\text{c}}^{\text{L}} - \eta_{\text{c}}^{\text{G}}$ in Fig. 9.b for a colloid-polymer mixture with $q=0.8$. The reported error bars are obtained by dividing the averaging run into 10 subaverages.  We find  qualitative agreement with results for the true binary mixture reported by Vink et al.~\cite{verso2006critical}

\section{\label{sec:conclusions}Conclusions}

In conclusion, we have introduced a coarse-graining ML approach for colloidal systems in which we trace out the degrees of freedom of the microscopic species and fit the resulting effective many-body potential  with a set of SFs using simple linear regression. We have applied this approach to a model suspension of colloidal particles and non-adsorbing polymer, and found that the ML potential accurately describes the effective many-body potential for a wide range of colloid densities and polymer fugacities. The ML potential can therefore be assumed to be effectively state-independent and can even be used to simulate direct phase coexistence. Given the computational efficiency of the fitted model, we were able to test the validity of the ML approach by measuring  the gas-liquid binodals using MC direct-coexistence simulations. We  found good agreement with previous results obtained by using simulations of the true binary mixture or using the full effective Hamiltonian that includes all many-body interactions, even for the largest $q$ that we studied, where four- and higher-body interactions are most pronounced. In addition, the structure was also well-captured by the ML potentials, but deviations appear when four- and higher-body contributions become predominant. Constructing higher-body encoders (SFs) seems to be necessary to describe more accurately the structure of the fluid. A generalization of the method to non-spherical bodies is important to account for the effective many-body interactions of anisotropic colloids, like mixtures  of non-adsorbing polymers with colloidal rods,\cite{patti2009multilayer,savenko2006phase} core-corona nanorods,\cite{campos2021} polyhedral-shaped particles,\cite{henzie2012self} and superballs.\cite{rossi2015shape} It will also be interesting to investigate whether this ML approach can  be extended to charged colloids, ligand-stabilized nanoparticles, starpolymers, which will be  subject of future work.

\begin{acknowledgments}
G.C.V. acknowledges funding from The Netherlands  Organisation  for  Scientific  Research  (NWO) for the ENW PPS Fund 2018 – Technology Area Soft Advanced Materials ENPPS.TA.018.002. M.D. has received funding from the European Research Council (ERC) under the European Union's Horizon 2020 research and innovation programme (Grant agreement No. ERC-2019-ADG 884902 SoftML). L.F. and E.B. acknowledge funding from The Netherlands  Organisation  for  Scientific  Research  (NWO)  (Grant  No. 16DDS004), and L.F. acknowledges funding from NWO for a Vidi grant (Grant No. VI.VIDI.192.102).
\end{acknowledgments}

\section*{Conflict of interest}
The authors have no conflicts to disclose.

\section*{Data Availability}
The data that support the findings of this study are available from the corresponding author upon reasonable request.

\appendix*

\section{Feature Selection Procedure }

In this appendix we compare the efficiency of the feature selection procedure used in this work (introduced in Ref.~\onlinecite{boattini2020modeling}) with two of the established selection methods by Imbalzano et al.~\cite{imbalzano2018} In particular, we test the Pearson Correlation method (PC), and the Farthest Point Sampling (FPS) scheme when using a linear regression scheme like we did in this work.

The PC and FPS methods introduced by Imbalzano et al. are based solely on the knowledge of the geometry of the particles' environments, and do not rely on the energy (or $V_{\text{f}}^{(3+)}$ in our case), nor on the performance of the model that results from a given choice of the SFs. Instead, the common idea behind those methods is to choose SFs which are as diverse as possible by, e.g., minimizing their linear correlation or maximizing their difference (in terms of their Euclidean distance), with the goal of minimizing the redundancy in the selected subset. The method we use, instead, aims at maximizing the correlation between the selected subset and the target $V_{\text{f}}^{(3+)}$  by iteratively selecting SFs that add relevant information (i.e. relevant to the task of fitting $V_{\text{f}}^{(3+)}$) to the previously selected set, while penalizing both (i) highly correlated SFs with only redundant information, and (ii) SFs which are sensitive to aspects of the particle' s environment that poorly correlate with $V_{\text{f}}^{(3+)}$. The inclusion of point (ii) is arguably the main difference with the other methods and guarantees a more efficient selection.

\begin{figure}[hbtp]
\includegraphics[scale=0.175]{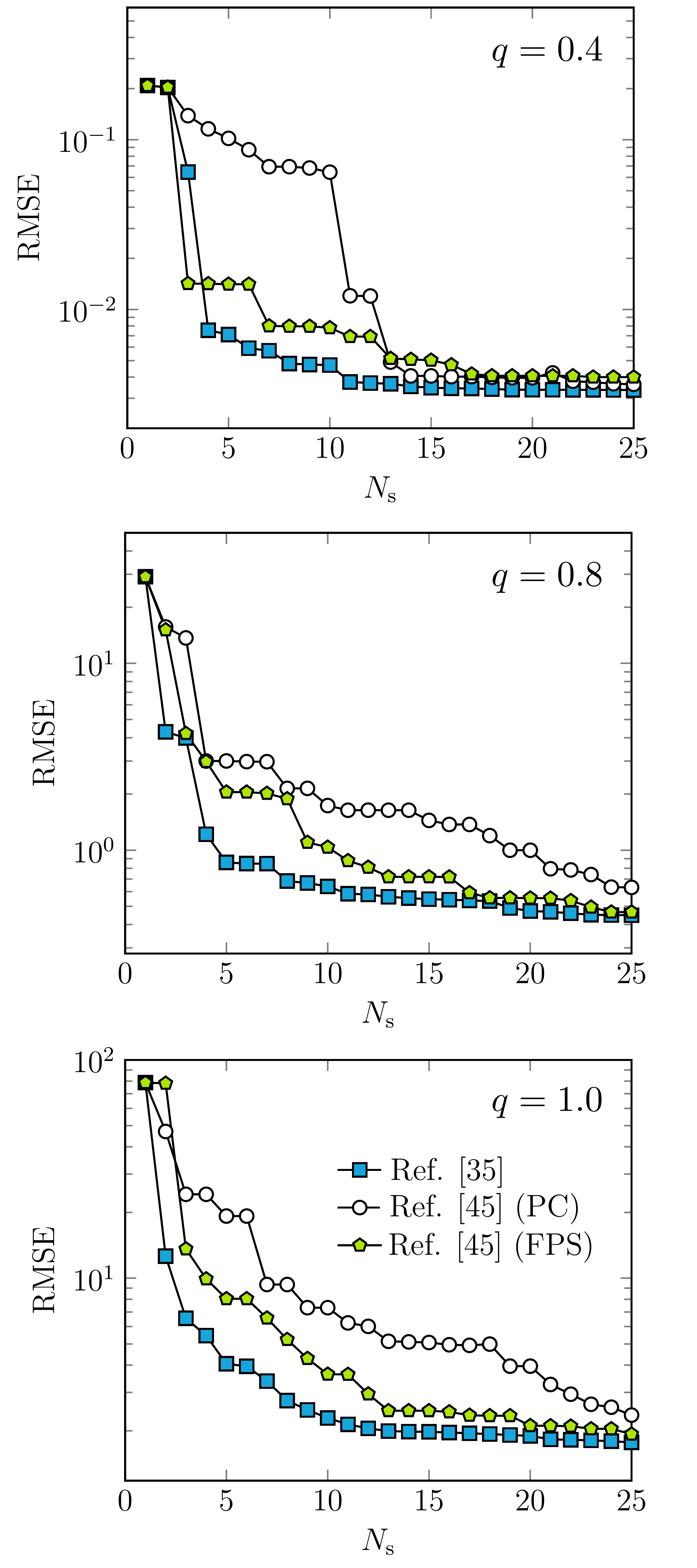}
\caption{\label{fig:met}Root mean squared error RMSE as a function of the number  of selected SFs $N_{\text{s}}$ for a colloid-polymer mixture with size ratio $q=0.4$, $0.8$ and $1.0$. We show the results for the selection procedure used in our work (Ref.~\onlinecite{boattini2020modeling}) and the PC and FPS schemes by Imbalzano et al.~\cite{imbalzano2018}}
\end{figure}

To demonstrate this, we show in Fig.~\ref{fig:met}, the performance of the three methods in terms of the RMSE on the test set as a function of the number of selected SFs for three different values of $q$. In all cases, the selection procedure of Ref. ~\onlinecite{boattini2020modeling} clearly outperforms the other methods, leading to a higher accuracy of the fit. Note that, in all cases and for all three methods, the first SF that is selected is the one with the largest correlation with $V_{\text{f}}^{(3+)}$.

\bibliography{aipsamp}

\end{document}